\pdfoutput=1
\documentclass[12pt]{article}
\usepackage[utf8]{inputenc}
\usepackage[height=8.85in,width=6.45in]{geometry}
\usepackage{hyperref}
\usepackage{todonotes}
\usepackage[most]{tcolorbox}

\usepackage{array}

\newtcolorbox[auto counter, number within=section]{statementbox}[1][]{
  colback=white, colframe=black,
  title=Statement~\thetcbcounter,
  fonttitle=\bfseries,
  boxrule=0.5pt,
  arc=2mm,
  left=6pt, right=6pt, top=6pt, bottom=6pt,
  #1
}

\newtcolorbox[auto counter, number within=section]{definitionbox}[1][]{
  colback=gray!10, colframe=gray!80,
  title=Definition~\thetcbcounter,
  fonttitle=\bfseries,
  boxrule=0.5pt,
  arc=2mm,
  left=6pt, right=6pt, top=6pt, bottom=6pt,
  #1
}

\newtcolorbox[auto counter, number within=section]{theorembox}[1][]{
  colback=gray!10, colframe=blue!80,
  title=Theorem~\thetcbcounter,
  fonttitle=\bfseries,
  boxrule=0.5pt,
  arc=2mm,
  left=6pt, right=6pt, top=6pt, bottom=6pt,
  #1
}

\newtcolorbox[auto counter, number within=section]{propositionbox}[1][]{
  colback=gray!10, colframe=green!60!black,
  title=Lemma~\thetcbcounter,
  fonttitle=\bfseries,
  boxrule=0.5pt,
  arc=2mm,
  left=6pt, right=6pt, top=6pt, bottom=6pt,
  #1
}

\newtcolorbox[auto counter, number within=section]{remarkbox}[1][]{
  colback=white, colframe=gray!50,
  title=Remark~\thetcbcounter,
  fonttitle=\bfseries,
  boxrule=0.3pt,
  arc=1.5mm,
  left=6pt, right=6pt, top=6pt, bottom=6pt,
  #1
}

\newtcolorbox{graybox}{breakable,
  colback=gray!40,     
  colframe=gray!15,    
  boxrule=0.2pt,       
  arc=0mm,             
  left=5pt, right=5pt, top=5pt, bottom=5pt
}

\newtcolorbox{lightgraybox}{breakable,
  colback=gray!7.5,     
  colframe=gray!15,    
  boxrule=0.2pt,       
  arc=0mm,             
  left=5pt, right=5pt, top=5pt, bottom=5pt
}

\usepackage{amsfonts}
\usepackage{amssymb}
\usepackage{amsmath}
\usepackage{amsthm}
\usepackage{mathtools}
\usepackage{cleveref}
\usepackage{url}
\usepackage{tikz}
\usepackage{tikz-cd}
\usepackage{adjustbox}
\usepackage{enumitem}
\usepackage{subcaption}
\usepackage{graphicx}
\usepackage{afterpage}
\usepackage{changepage}
\usepackage{authblk}
\usepackage{longtable}
\usepackage{bm}
\usepackage{multicol}
\usepackage{wrapfig}
\usepackage{enumitem}
\usepackage{braket}

\usetikzlibrary{knots}

\theoremstyle{plain}

\theoremstyle{definition}

\theoremstyle{remark}

\crefname{maintheorem}{Theorem}{Theorems}
\crefname{claim}{Claim}{Claims}

\newtheoremstyle{restated}
    {\topsep}{\topsep} 
    {\itshape}         
    {}                 
    {\bfseries}        
    {.}                
    { }                
    {\thmname{#1} \ref{#3} {\normalfont(Restated)}}
    
\theoremstyle{restated}
    \newtheorem{restate-theorem}{Theorem}
    \newtheorem{restate-proposition}{Proposition}
    \newtheorem{restate-corollary}{Corollary}
    
\numberwithin{equation}{section}
\numberwithin{table}{section}

\renewcommand{\epsilon}{\varepsilon}

\DeclareMathOperator{\Ima}{Im}
\DeclareMathOperator{\Ker}{Ker}

\newcommand{\sT}{\mathsf{T}}
\newcommand{\sG}{\mathsf{G}}

\newcommand{\mT}{\mathbf{T}}

\newcommand{\bZ}{\mathbb{Z}}

\newcommand{\cA}{\mathcal{A}}
\newcommand{\cC}{\mathcal{C}}

\title{Vanishing of the $H^3$ obstruction
for time-reversal symmetry in (2+1)D abelian bosonic TQFTs}

\author{Ippo Orii}

\affil{Kavli Institute for the Physics and Mathematics of the Universe, \\
University of Tokyo, Kashiwa, Chiba 277-8583, Japan}

\date{}
\begin{document}
    \maketitle
    
    \begin{abstract}
In $(2+1)$-dimensional topological quantum field theories (TQFTs), the action of a global symmetry group on the anyon system is one of the central topics of research. 
Owing to the subtle categorical nature of anyons, a global symmetry acting on them is generally realized in a projective manner. Symmetry fractionalization encodes this projective realization. The obstruction to defining symmetry fractionalization is captured by a cohomology class, known as the $H^3$ obstruction, whose nontriviality signals a failure to define symmetry fractionalization consistently. In this short note, we prove that the $H^3$ obstruction for time-reversal symmetry always vanishes in abelian bosonic TQFTs.

\end{abstract}

    \setcounter{tocdepth}{2}
    \tableofcontents
    \clearpage

    \section{Introduction and Summary}
Topological quantum field theory (TQFT) is a quantum field theory that does not depend on the background metric. Since its introduction in the early 1990s, TQFT has been extensively studied from both physical and mathematical perspectives. In $(2+1)$ dimensions, TQFTs exhibit particularly rich structures through the appearance of anyons, which are naturally described by modular tensor categories $\cC$.

One central theme in the study of $(2+1)$D TQFTs is the action of global symmetry groups on anyons~\cite{Barkeshli_2019}. To encode a  global unitary action of a group $G$ on an anyon system $\cC$, one attempts to construct a $G$-crossed modular tensor category
\begin{equation}
       \cC_G^\times = \bigoplus_{g\in G} \cC_g,
    \qquad \text{with } \cC_{\mathrm{id}} = \cC.
\end{equation}
Here, $\cC_{\mathrm{id}}$ describes the original anyon system, and $\cC_g$ describes the defects labeled by group elements $g\in G$. Associativity in this setting takes the form
\begin{equation}
     a_g \otimes (b_h \otimes c_k) \;\cong\; (a_g \otimes b_h) \otimes c_k,
\end{equation}
where $a_g\in\cC_g$, $b_h\in\cC_h$, and $c_k\in\cC_k$. However, in the $G$-crossed case, associativity does not necessarily hold strictly. Instead, there may exist an invertible anyon $e(g,h,k)\in \cA$ such that
\begin{equation}
     a_g \otimes (b_h \otimes c_k)
    \;\cong\; \bigl((a_g \otimes b_h) \otimes c_k\bigr) \otimes e(g,h,k),
\end{equation}
where $\cA \subset \cC$ denotes the subgroup of invertible anyons. The data $e$ defines a class in twisted group cohomology
\begin{equation}
       e \in H^3_{[\rho]}(G,\cA),
\end{equation}
which is referred to in the physics literature as the \emph{symmetry localization obstruction} and in the mathematics literature as the \emph{Postnikov class}. We will simply call it the $H^3$ obstruction. This class obstructs strict associativity of the group action on the anyon system. In the absence of this obstruction, symmetry fractionalization is well defined. It encodes the projective phases associated with the symmetry action on anyons and is classified by a torsor over \(H^2_{[\rho]}(G,\cA)\), which must be specified to uniquely determine the theory.

When the global symmetry is time reversal, i.e., an anti-unitary \(\bZ_2\) action, there is no known notion of a \(G\)-crossed extension. In this sense, the obstruction does not admit the interpretation described above. Nevertheless, it is known that it can still be understood as an obstruction to defining symmetry fractionalization~\cite{Barkeshli:2016mew}, and  described in terms of 2-group~\cite{Benini_2019}.

In the unitary case, for finite symmetry groups \(G\), no example of an abelian system with a nontrivial \(H^3\) obstruction is currently known, although no general proof of its triviality exists.\footnote{Assuming anomaly inflow, the obstruction is shown to be trivial for all abelian TQFTs; see the footnote in~\cite[Sec.~5.3]{Benini_2019}.}

In time-reversal cases,
several necessary conditions for the vanishing of the obstruction—valid for both abelian and non-abelian systems—were proposed in~\cite{Barkeshli:2017rzd}.\footnote{Note that they also put an assumption in terms of anomaly inflow, which is very likely for physicists ourselves.} In the abelian case, we verified in~\cite{Orii_2025} that these conditions are always satisfied. Furthermore, the obstruction was shown to vanish when $|\cA|$ is odd in~\cite{Cui:2015cbf,etingof2018reflectionfusioncategories}. However,  there has been no general proof that the obstruction vanishes for all abelian TQFTs. 

In this short note, we provide a proof that the obstruction for an unitary/anti-unitary \(\bZ_2\) symmetry is always trivial, without assuming anomaly inflow.

\paragraph{Organization of the paper :} 
In Sec.~2 we give a brief review of abelian TQFTs from several perspectives. 
In Sec.~3, we describe the time-reversal action on anyon systems and define the $H^3$ obstruction. 
In Sec.~4, we prove that the $H^3$ obstruction vanishes for time-reversal symmetry and also comment on the unitary $\bZ_2$ case.

    \section{Basics of abelian TQFTs}
In this note, we employ an abstract formalism to describe our theories. 
Nevertheless, all of our constructions can also be formulated explicitly in the Lagrangian framework. 
To set the stage, we begin by reviewing $U(1)^N$ Chern--Simons theory, the prototypical example of an abelian TQFT. 
We then introduce a minimal abstract framework for abelian TQFTs. 
Finally, we present an alternative description based on Moore--Seiberg data, and comment on the equivalence among these three constructions.

\subsection{$U(1)^N$ Chern--Simons Theory}
A general abelian Chern--Simons theory with gauge group $U(1)^N$ is specified by the choice of an
integral symmetric matrix $K$ of size $N \times N$, known as the \emph{$K$-matrix}.
The defining properties of $K$ are
\begin{itemize}
    \item $K$ is symmetric,
    \item $K$ has integral entries,
    \item $K$ has even diagonal entries (this condition is required if we restrict to purely bosonic theories).
\end{itemize}

The corresponding Lagrangian takes the form
\begin{equation}
    \label{eq:CS-Lagrangian}
    \mathcal{L} = \frac{1}{4\pi} K_{IJ}\, a_I \wedge d a_J ,
\end{equation}
where $\{a_I\}_{I=1}^N$ are dynamical $U(1)$ gauge fields.

\medskip

Anyons of the theory are labeled by integer vectors modulo the $K$-lattice,
\begin{equation}
    a \in \mathbb{Z}^N / K \mathbb{Z}^N .
\end{equation}
It is also worth noting that the number of distinct anyon types is determined directly by the $K$-matrix:
\begin{equation}
    \bigl|\mathbb{Z}^N / K \mathbb{Z}^N \bigr|
    \;=\; \bigl|\det K \bigr| .
\end{equation}
Fusion is simply given by addition in this quotient group,
\begin{equation}
  a+b \quad \in \mathbb{Z}^N / K \mathbb{Z}^N .
\end{equation}

\medskip

Two key topological observables are the topological spin and the mutual braiding phase. 
They are expressed as
\begin{align}
    \theta(a) &= \exp\!\left( \pi i\, a^{T} K^{-1} a \right),\\
    B(a,b) &= \exp\!\left( 2\pi i\, a^{T} K^{-1} b \right),
\end{align}
for $a,b \in \mathbb{Z}^N / K \mathbb{Z}^N$.
These expressions are well-defined on the quotient: under the replacement
$a \mapsto a + K\lambda$ with $\lambda \in \mathbb{Z}^N$, they remain invariant due to integrality and symmetry of $K$.

\medskip

Thus, the data of the $K$-matrix encodes the full abelian anyon content of the theory, including
fusion rules, spins, and braiding phases.

\medskip

Finally, let us mention an important identity, sometimes referred to as the Gauss--Milgram formula (see~\cite{Belov:2005ze}):
\begin{equation}
\frac{1}{\left|\det K\right|^{1/2}} 
   \sum_{a \in \mathbb{Z}^N / K \mathbb{Z}^N} 
   \exp\!\left( \pi i\, a^{T} K^{-1} a \right) 
   \;=\; 
   \exp\!\left( \tfrac{2\pi i}{8}\,\mathrm{sgn}(K) \right).
\end{equation}
Here $\mathrm{sgn}(K)$ denotes the signature of the matrix $K$, i.e.\ the number of positive eigenvalues minus the number of negative eigenvalues. 
This identity plays a crucial role, as it implies that the data of the abelian anyon theory encodes the chiral central charge, which we shall define and discuss in the next subsection.

\subsection{Defining data of abelian systems}
Let us define the minimal data required for an abelian anyon system. Here, we consider abelian \emph{bosonic} systems, which are well-defined without specifying a spin structure on the spacetime manifold. The required data are as follows:
\begin{lightgraybox}
    \begin{itemize}
  \item \( \mathcal{A} \): a finite abelian group\footnotemark{} of anyons (i.e., the group of topological charges).
  \item \( \theta \): the topological spin, a function \( \theta \colon \mathcal{A} \to U(1) \) which is non-degenerate, quadratic, and homogeneous.
  \item \( c_- \): the chiral central charge, an integer \( c_- \in \mathbb{Z} \) satisfying the Gauss sum constraint.
\end{itemize}
\end{lightgraybox}

\footnotetext{In general, the fusion rules of anyons are defined through the decomposition of the tensor product into a direct sum of simple objects. A theory is called \emph{Abelian} if the fusion of any two simple anyons results in another simple anyon—that is, the tensor product does not decompose further. In such cases, we use additive notation: for example, we write \( a + b \coloneqq a \otimes b \) and \( a - b \coloneqq a \otimes \overline{b} \), where \( \overline{b} \) denotes the antiparticle of \( b \).}

Given the data above, we define the braiding phase by:
\begin{equation}
    \begin{array}{rccc}
        B\colon & \mathcal{A} \times \mathcal{A} & \longrightarrow & U(1) \\
               & \rotatebox{90}{$\in$} & & \rotatebox{90}{$\in$} \\
               & (a, b) & \longmapsto & \theta(a + b)\theta(a)^{-1}\theta(b)^{-1}
    \end{array}.
\end{equation}

We use the following terminology:
\begin{itemize}
  \item \( \theta \) is called \emph{non-degenerate} if the associated braiding \( B \) is a non-degenerate pairing.
  \item \( \theta \) is called \emph{quadratic} if \( B \) is bihomomorphic.
  \item \( \theta \) is called \emph{homogeneous} if
  \begin{equation}
      \theta(n a) = \theta(a)^{n^2} \quad \text{for all } a \in \mathcal{A},\, n \in \mathbb{Z}.
  \end{equation}
\end{itemize}

Finally, the Gauss sum constraint is given by:
\begin{equation}
    \frac{1}{|\mathcal{A}|^{1/2}} \sum_{a \in \mathcal{A}} \theta(a) = e^{2\pi i c_- / 8}.
\end{equation}

\paragraph{Relation to $U(1)^N$ Chern--Simons theory :}It is known that two distinct Lagrangian descriptions of $U(1)^N$ Chern--Simons theory that yield the same triple $(\cA,\theta,c)$ are equivalent~\cite{Belov:2005ze}. Conversely, any such triple $(\cA,\theta,c)$ can be realized from an even integral lattice $(\mathbb{Z}^N,K_{IJ})$~\cite{WALL1963281,Nikulin1980}. In this sense, every abelian TQFT arises from some $U(1)^N$ Chern--Simons theory. Although we adopt an abstract formalism to describe TQFTs, the description can always be translated into a Lagrangian framework.

\subsection{Moore--Seiberg Data}
To carry out computations within the framework of a $(2+1)$-dimensional TQFT, 
one requires the \emph{Moore--Seiberg data}~\cite{Moore:1988qv,Moore:1989vd}, 
or equivalently, a description of anyons in terms of a modular tensor category. 
For completeness, we briefly recall the essential features below.
\vspace{1em}

To define the fusion and braiding of  abelian anyons, we define the following maps:
\begin{lightgraybox}
\textbf{$F$-symbols and $R$-symbols :}
\begin{align}
  F&:\cA\times\cA\times\cA\to U(1)\\
  R&:\cA\times\cA\to U(1)
\end{align}
satisfying the following relations:\\
\textbf{Pentagon relation :} 
    \begin{equation}
          F(a, b, c + d)F(a + b, c, d) = F(b, c, d)F(a, b + c, d)F(a, b, c).
    \end{equation}
    \textbf{Hexagon relation :} 
    \begin{align}
    R(a, b + c) &= F(a, b, c)^{-1} R(a, b) F(b, a, c) R(a, c) F(b, c, a)^{-1}, \\
    R(a + b, c) &= F(a, b, c) R(b, c) F(a, c, b)^{-1} R(a, c) F(c, a, b).
\end{align}
\end{lightgraybox}
\medskip

In terms of the braiding $R$, the topological spin and mutual braiding can be expressed as
\begin{equation}
    \theta(a) = R(a,a), 
    \qquad 
    B(a,b) = R(b,a)R(a,b).
\end{equation}

\medskip

It is important to note that there remains a redundancy in the choice of $F$- and $R$-symbols, which may be regarded as a kind of gauge freedom. 
For a function $U : \cA \times \cA \to U(1)$, we can define transformed versions of the associator $F$ and braiding $R$ by
\begin{align}
    (U.F)(a,b,c) &:= U(b,c)\, U(a+b,c)^{-1}\, U(a,b+c)\, U(a,b)^{-1}\, F(a,b,c), \\
    (U.R)(a,b) &:= U(a,b)^{-1}\, U(b,a)\, R(a,b).
\end{align}
One can check that $(U.F, U.R)$ continue to satisfy the Pentagon and Hexagon equations. 
Moreover, they leave $\theta$ and $B$ invariant, so the two sets of data $(F,R)$ and $(U.F, U.R)$ describe the same physical theory.

\medskip

A natural question is: when do two pairs $(F,R)$ and $(U.F,U.R)$ define \emph{exactly} the same structure, i.e. $(F,R)=(U.F,U.R)$.
It is known that this occurs if and only if the transformation $U$ arises from a map $\beta : \cA \to U(1)$ via
\begin{equation}
    U(a,b) = \frac{\beta(a)\beta(b)}{\beta(a+b)}.
\end{equation}

\paragraph{Comment on three constructions :}
So far, we briefly introduced three constructions of abelian TQFTs. 
In the previous subsection, we reviewed that the first two constructions are equivalent. 
Furthermore, the third construction, described by the pair $(F,R)$, is in one-to-one correspondence with the other two. 
For a more detailed discussion, see~\cite{Lee:2018eqa}.

    \section{What is the obstruction}

Group actions on general, possibly non-abelian anyons were studied in detail in~\cite{Barkeshli_2019} from both mathematical and condensed matter perspectives. The resulting equations are often quite cumbersome. In this section, we restrict our attention to the action of time-reversal symmetry on abelian anyons, which simplifies our discussion. This section is a review of discussions in~\cite{Lee:2018eqa}.

\subsection{Time Reversal on Moore--Seiberg Data}

\paragraph{Time reversal on anyon :}
First, let us introduce a time-reversal action on anyons \( \mathsf{T} \colon \mathcal{A} \to \mathcal{A} \) by the condition:
\begin{lightgraybox}
  \begin{equation}
    \mathsf{T} \colon \mathcal{A} \to \mathcal{A}, \quad \text{such that }\sT^2=\mathrm{id}_\cA,\quad \theta(\mathsf{T}a)\, \theta(a) = 1 \quad \text{for all } a \in \mathcal{A}.
\end{equation}  
\end{lightgraybox}
The condition \( \theta(\mathsf{T}a)\, \theta(a) = 1 \) captures the anti-unitary nature of time-reversal symmetry (see, e.g., \cite{Barkeshli:2016mew}), and we will sometimes write \(\sT^2=\mathrm{id}_\cA\) compactly as \( \mathsf{T}^2 = 1 \).\footnote{Note that time-reversal symmetry can be encoded in the $K$-matrix formalism by an integer matrix $\sT$ satisfying
\(\sT^{T} K \sT = -K\).
For example, one can check that
\(\theta(\sT a)
= \exp\!\bigl(\pi i (\sT a)^{T} K^{-1} (\sT a)\bigr)
= \overline{\theta(a)}\),
which is the required condition for the action of time-reversal symmetry on anyons.
}


\paragraph{Time reversal on $(F,R)$ :}
We fix the Moore--Seiberg data \((F, R)\) associated with the anyon system \((\cA, \theta)\). The time-reversed Moore--Seiberg data \((\sT F, \sT R)\) is defined by
\begin{equation}
    \sT F(a, b, c) :=\overline{F(\sT a, \sT b,\sT c)}, \qquad \sT R(a, b) := \overline{R(\sT a, \sT b)}.
\end{equation}
This pair \((\sT F, \sT R)\) also satisfies the Pentagon and Hexagon relations and gives us the same value of $\theta$ and $B$. Therefore, there exist phases \(U(a, b) \in U(1)\) such that
\begin{equation}
    (\sT F, \sT R)= (U.F, U.R),\label{(TFR,UFR)}
\end{equation}
where the right-hand side denotes a gauge transformation as defined previously.

\medskip

Since \(\sT^2 = \mathrm{id}\), applying time reversal twice must return the original data, i.e., \((F, R) = (\sT\sT F,\sT\sT R)\). Applying the gauge transformation twice yields
\begin{equation}
    (F, R) = (\kappa.F, \kappa.R), \qquad \text{where} \quad \kappa(a, b) := \overline{U(\sT a, \sT b)}\, U(a, b).
\end{equation}
This implies that there exists a map \(\beta \colon A \to U(1)\) such that
\begin{equation}
    \overline{U(\sT a, \sT b)}\,U(a, b) = \frac{\beta(a)\beta(b)}{\beta(a + b)}.\label{U and beta}
\end{equation}

\vspace{1em}

Now, for such a map $\beta$, we define another phase $\Omega(a)$ by
\begin{lightgraybox}
\begin{equation}
     \Omega(a) := \frac{1}{\beta(\sT a)\beta(a)}.
\end{equation}
\end{lightgraybox}
It is straightforward to verify that
\begin{equation}
    \Omega(a) = \Omega(\sT a), \label{omega a and ta}
\end{equation}
and
\begin{equation}
    \Omega(a+b) = \Omega(a)\Omega(b), \label{linear omega}
\end{equation}
for all $a,b \in \cA$.

\paragraph{Redundancy for $U$ :} 
If $U$ satisfies Equation~\eqref{(TFR,UFR)}, then so does $\hat{U}$ defined by
\begin{equation}
     \hat{U}(a,b) := U(a,b)\,\frac{\gamma(a)\gamma(b)}{\gamma(a+b)},
\end{equation}
for any map $\gamma : \cA \to U(1)$. 
Under this redefinition, $\beta$ is shifted but $\Omega$ remains invariant:
\begin{equation}
       \hat{\beta}(a) := \beta(a)\,\overline{\gamma(\sT a)}\,\gamma(a),
    \qquad 
    \hat{\Omega}(a) = \Omega(a), \quad \text{for all } a \in \cA.
\end{equation}

\paragraph{Redundancy for $\beta$ :}
Similarly, if $\beta$ satisfies Equation~\eqref{U and beta}, then so does
\begin{equation}
     \Tilde{\beta}(a) := \beta(a)\nu(a),
\end{equation}
where $\nu:\cA \to U(1)$ is a homomorphism, i.e.
\begin{equation}
     \nu(a+b) = \nu(a)\nu(b).
\end{equation}
In this case, $\Omega$ transforms as
\begin{equation}
    \Tilde{\Omega}(a) = \Omega(a)\,\frac{1}{\nu(\sT a)\nu(a)}. \label{tilda omega}
\end{equation}
The additional phase factor should be interpreted as 
physically trivial, in the sense that it does not affect Equation~\eqref{U and beta}.

\subsection{Definition of the $H^3$ obstruction}
We are now ready to define the obstruction referred to in the title of this note. 
Let us begin with the following fact:

\begin{lightgraybox}
    If $f:\cA \to U(1)$ satisfies
    \begin{equation}
          f(a+b) = f(a)f(b) \qquad \text{for all } a,b \in \cA,
    \end{equation}
    then there exists an element $\boldsymbol{f} \in \cA$ such that
    \begin{equation}
         f(a) = B(a,\boldsymbol{f}) \qquad \text{for all } a \in \cA.
    \end{equation}
\end{lightgraybox}

Using this fact, we can represent $\Omega$ as
\begin{equation}
    \Omega(a) = B(a,\boldsymbol{\Omega}) \qquad \text{for all } a \in \cA,
\end{equation}
for some $\boldsymbol{\Omega} \in \cA$, since $\Omega$ satisfies~\eqref{linear omega}. 
From this representation one finds
\begin{graybox}
    \begin{equation}
            \boldsymbol{\Omega} \in \Ker(1+\sT).
    \end{equation}
\end{graybox}

Indeed, by combining~\eqref{omega a and ta} and~\eqref{linear omega}, we obtain
\begin{equation}
      \Omega\bigl((1-\sT)a\bigr)
      = \Omega(a)\Omega(-\sT a)
      = \frac{\Omega(a)}{\Omega(\sT a)}
      = 1
      \qquad \text{for all } a \in \cA.
\end{equation}
By the definition of $\boldsymbol{\Omega}$, this implies
\begin{equation}
    \boldsymbol{\Omega} \in \Ima(1-\sT)^\perp = \Ker(1+\sT),
\end{equation}
where in the last equality we have used the property reviewed in Appendix~\ref{App 1}.

\medskip

On the other hand, recalling the definition of $\Tilde{\Omega}$ and $\nu$, we have
\begin{equation}
     \boldsymbol{\Tilde{\Omega}} = \boldsymbol{\Omega} - (1-\sT)\boldsymbol{\nu},
\end{equation}
where $\boldsymbol{\Tilde{\Omega}}$ and $\boldsymbol{\nu}$ are defined by
\begin{equation}
      \Tilde{\Omega}(a) = B(a,\boldsymbol{\Tilde{\Omega}}),
    \qquad
    \nu(a) = B(a,\boldsymbol{\nu})
    \qquad \text{for all } a \in \cA.
\end{equation}

Since $\Tilde{\Omega}$ should be regarded as physically equivalent to $\Omega$, the relevant data is not $\boldsymbol{\Omega}$ itself but rather its equivalence class:
\begin{graybox}
    \begin{equation}
         [\boldsymbol{\Omega}] \;\in\; \frac{\Ker(1+\sT)}{\Ima(1-\sT)}.
    \end{equation}
\end{graybox}
This equivalence class is what we call the $H^3$ obstruction. 
This element \(\boldsymbol{\Omega}\) introduced above arises from a third group cohomology class. Let us briefly explain this point.
We denote \(\mathbb{Z}_2 = \{1,\mathsf{T}\}\) and consider the twisted group cochain complex
\begin{equation}
      d^{(p)} : C_{[\rho]}^p(\mathbb{Z}_2,\cA) \to  C_{[\rho]}^{p+1}(\mathbb{Z}_2,\cA).
\end{equation}
Suppose $a \in Z_{[\rho]}^3(\mathbb{Z}_2,\cA) := \Ker d^{(3)}$ is a 3-cocycle. 
Then one computes\footnote{Here, we write \(\sT x:=\rho_{\mT} (x)\) for \(x\in\cA\).}\footnote{We set the normalization condition here, i.e. \(a(g_1,g_2,g_3)=0\quad\text{if }g_i=1\quad\text{for some }i\in\{1,2,3\}.\) See e.g.~\cite[Chapter~IV]{MacLane1995Homology}}
\begin{equation}
    \begin{aligned}
          d^{(3)} a(\mT,\mT,\mT,\mT)
      &= \sT a(\mT,\mT,\mT) - a(1,\mT,\mT) + a(\mT,1,\mT) - a(\mT,\mT,1) + a(\mT,\mT,\mT) \\
      &= (1+\sT)\,a(\mT,\mT,\mT) \\
      &= 0.
    \end{aligned}
\end{equation}
Therefore
\begin{equation}
     a(\mT,\mT,\mT) \;\in\; \Ker(1+\sT).
\end{equation}

Likewise, for a 2-cochain $b$, one has $d^{(2)} b(\mT,\mT,\mT) \in \Ima(1-\sT)$. 
Consequently, if $\omega \in H_{[\rho]}^3(\mathbb{Z}_2,\cA) = \Ker d^{(3)} / \Ima d^{(2)}$, then
\begin{equation}
        \omega(\mT,\mT,\mT) \;\in\; \frac{\Ker(1+\sT)}{\Ima(1-\sT)}.
\end{equation}

$\boldsymbol{\Omega}$ is the element of this class:
\begin{equation}
    \boldsymbol{\Omega}\in\frac{\Ker(1+\sT)}{\Ima(1-\sT)}.
\end{equation}

\paragraph{Symmetry fractionalization.}
When the obstruction vanishes, we can choose \(\beta(a)\) such that \(\boldsymbol{\Omega}=0\in\cA\).
We denote such a choice of \(\beta(a)\) by \(\eta(a)\).
In other words, \(\eta(a)\) satisfies
\begin{equation}
     \overline{U(\sT a, \sT b)}\,U(a, b)
     = \frac{\eta(a)\eta(b)}{\eta(a + b)},
     \qquad
     \eta(\sT a)\,\eta(a)=1.
\end{equation}
This choice of \(\eta(a)\) defines the symmetry fractionalization for time-reversal symmetry.
From these equations, one can derive the general properties of symmetry fractionalization, such as its torsor structure.
See~\cite[Sec.~2.4]{Lee:2018eqa}.

    \section{Proof of the vanishing of the $H^3$ obstruction}
\subsection{$H^3$ obstruction for time reversal}
In this section, we prove that the obstruction is in fact trivial:
\begin{graybox}
\begin{center}
   $H^3$ obstruction for time reversal vanishes, i.e. \(\boldsymbol{\Omega}\in \Ima(1-\sT)\).
\end{center}
\end{graybox}

By the orthogonality relation 
\(\Ima(1-\sT)^\perp = \Ker(1+\sT)\), 
it suffices to show that
\begin{equation}
    B(a,\boldsymbol{\Omega}) = 1
    \quad \text{for all } a \in \Ker(1+\sT).
\end{equation}

\medskip

From the definitions of $\Omega$ and $\kappa$, we have
\begin{equation}
    \begin{aligned}
        B(a,\boldsymbol{\Omega})
        &= \Omega(a) \\
        &= \frac{1}{\beta(\sT a)\beta(a)} \\
        &= \frac{1}{\beta(-a)\beta(a)} \\
        &= \frac{1}{\kappa(a,-a)},
    \end{aligned}
\end{equation}
where we used $a \in \Ker(1+\sT)$, i.e.\ $a = -\sT a$. 
We also note that $\beta(0)=1$, since
\begin{equation}
    \begin{aligned}
        \beta(0)
      &= \frac{\beta(0)\beta(0)}{\beta(0+0)} \\
      &= \overline{U(0,0)}\,U(0,0) \\
      &= 1.
    \end{aligned}
\end{equation}

\medskip

Next, $\kappa(a,-a)$ can be rewritten as
\begin{equation}
    \begin{aligned}
         \kappa(a,-a)
      &= \overline{U(\sT a,-\sT a)}\,U(a,-a) \\
      &= \overline{U(-a,a)}\,U(a,-a) \\
      &= U(-a,a)^{-1}\,U(a,-a).
    \end{aligned}
\end{equation}
Recalling the relation
\begin{equation}
    \sT R(a,b)
      = \overline{R(\sT a,\sT b)}
      = U(a,b)^{-1}\,U(b,a)\,R(a,b),
\end{equation}
we find
\begin{equation}
    \begin{aligned}
         U(-a,a)^{-1}U(a,-a)
      &= R(a,-a)R(\sT a,-\sT a) \\
      &= R(a,-a)R(-a,a) \\
      &= B(-a,a).
    \end{aligned}
\end{equation}
Using the definition of the braiding $B$, this becomes
\begin{equation}
    \begin{aligned}
         B(-a,a)
      &= \frac{\theta(-a+a)}{\theta(-a)\theta(a)} \\
      &= \frac{1}{\theta(\sT a)\theta(a)} \\
      &= 1.
    \end{aligned}
\end{equation}

\medskip

Combining these computations, we obtain
\begin{equation}
     B(a,\boldsymbol{\Omega}) = 1
    \quad \text{for all } a \in \Ker(1+\sT),
\end{equation}
which implies
\begin{equation}
    \boldsymbol{\Omega} \in \Ima(1-\sT).
\end{equation}

\medskip

This completes the proof of triviality of the obstruction.

\subsection{Comment on the unitary case}

We have proved that the $H^3$ obstruction vanishes for time-reversal symmetry, i.e.\ for an anti-unitary $\bZ_2$ action. The same conclusion also holds for a unitary $\bZ_2$ symmetry. Since the proof essentially follows by imitating the time-reversal case---and is in fact simpler---we only provide a rough sketch here.

\medskip

The starting point is as follows:
\begin{lightgraybox}
    \paragraph{Unitary $\bZ_2$ on anyons :} Define $\sG:\cA\to\cA$ by
    \begin{equation}
         \sG:\cA\to\cA,\qquad \sG^2=\mathrm{id}_\cA, \quad \theta(\sG a)=\theta(a)\quad\text{for all }a\in\cA.
    \end{equation}
    \paragraph{Unitary $\bZ_2$ on $(F,R)$ :} Define $(\sG F,\sG R)$ by
    \begin{equation}
         \sG F(a,b,c):=F(\sG a,\sG b,\sG c),\qquad \sG R(a,b):=R(\sG a,\sG b).
    \end{equation}
    \paragraph{Definitions of $U$, $\beta$, and $\Omega$ :} Define $U$, $\beta$, and $\Omega$ by
    \begin{align}
        (\sG F,\sG R)&=(U.F,U.R),\\
        U(\sG a,\sG b)U(a,b)&=\frac{\beta(a)\beta(b)}{\beta(a+b)},\\
        \Omega(a)&=\frac{\beta(\sG a)}{\beta(a)}.
    \end{align}
\end{lightgraybox}

The difference from the anti-unitary case comes only from the absence of complex conjugation~\cite{Barkeshli_2019}. In this setup, the obstruction is again defined as
\begin{equation}
    \boldsymbol{\Omega}\in \frac{\Ker(1+\sG)}{\Ima(1-\sG)}.
\end{equation}
From the definition of the $\sG$-action on $\theta$, one obtains orthogonality relations analogous to Appendix~\ref{App 1}:
\begin{graybox}
\begin{equation}
        \Ker(1 - \sG) = \left[ \Ima(1 - \sG) \right]^\perp, \qquad 
    \Ker(1 + \sG) = \left[ \Ima(1 + \sG) \right]^\perp.
\end{equation}
\end{graybox}
Using these relations, it is straightforward to check that
\begin{graybox}
    \begin{equation}
         \boldsymbol{\Omega}\in\Ima(1-\sG).
    \end{equation}
\end{graybox}
Therefore, we conclude that the $H^3$ obstruction vanishes in abelian TQFTs for both unitary and anti-unitary $\bZ_2$ symmetries, although the main focus of this paper is on the anti-unitary case.
\paragraph{Interpretation in the unitary case :}When the action of the group is unitary, the \(H^3\) obstruction admits a more concrete interpretation and is not merely an obstruction to defining symmetry fractionalization.
For the reader’s convenience, let us explain this in such a setup.

\medskip

Let us now consider a finite group $G$ acting unitarily on an anyon system $\cC$, not necessarily an abelian system. 
In this situation, one attempts to construct a $G$-crossed modular tensor category
\begin{equation}
     \cC_G^\times = \bigoplus_{g\in G} \cC_g,
    \qquad \text{with } \cC_{\mathrm{id}} = \cC.
\end{equation}
We will not enter into the full details of this construction (see~\cite{Cui:2015cbf, Barkeshli_2019} for general discussions). 
Roughly speaking, an object $a_g \in \cC_g$ corresponds to a non-genuine line which is a boundary of a branch sheet labeled by $g \in G$, 
and anyons passing through the branch sheet are acted on by $g$. 
These branch-sheet labels behave like group elements under fusion:
\begin{equation}
    a_g \otimes b_h \;\in\; \cC_{gh},
\end{equation}
with the fusion satisfying the associativity constraint
\begin{equation}
      a_g \otimes (b_h \otimes c_k) \;\cong\; (a_g \otimes b_h) \otimes c_k .
\end{equation}
However, in the $G$-crossed setting, associativity no longer holds trivially. 
There may exist an invertible anyon $e(g,h,k) \in \cA$ such that
\begin{equation}
      a_g \otimes (b_h \otimes c_k)
    \;\cong\; \bigl((a_g \otimes b_h) \otimes c_k\bigr) \otimes e(g,h,k),
\end{equation}
where $\cA \subset \cC$ denotes the subgroup of invertible anyons. Here we encounter precisely the obstruction mentioned above. 
In this setting, the element $e$ defines a class in a twisted third group cohomology\footnote{The index $[\rho]$ encodes the way in which the group $G$ permutes the anyons. 
More precisely, a group $G$ acts on an anyon system via a group homomorphism
\begin{equation}
      [\rho] : G \to \mathrm{Aut}(\cC),
\end{equation}
where $\mathrm{Aut}(\cC)$ denotes the group of braided tensor autoequivalences of $\cC$, taken up to natural isomorphism. 
For further details, see~\cite{Barkeshli_2019}.
}
\begin{equation}
    e \in H^3_{[\rho]}(G,\cA),
\end{equation}
which is referred to in the physics literature as the \emph{symmetry localization obstruction}, 
and in the mathematics literature as the \emph{Postnikov class}. 
This class obstructs the strict associativity of the group action on the anyon system. 
From the modern perspective, this is an example of a mixed symmetry structure 
involving both $0$-form and $1$-form symmetries, 
and is naturally captured by the language of $2$-groups~\cite{Tachikawa_2020,Benini_2019}.

\paragraph{Acknowledgements :} 
The author would like to thank Y. Tachikawa for helpful discussions. He would also like to thank the anonymous referee for their helpful comments.
 
This work was supported in part by the Forefront Physics and Mathematics Program to Drive Transformation (FoPM), a World-leading Innovative Graduate Study (WINGS) program at the University of Tokyo.

    \appendix
\section{Orthogonality of $\Ker(1-\sT)$ and $\Ima(1+\sT)$}\label{App 1}
In this section, we will prove the following equalities:
\begin{graybox}
    \begin{equation}
    \Ker(1 - \sT) = \left[ \Ima(1 + \sT) \right]^\perp, \quad 
    \Ker(1 + \sT) = \left[ \Ima(1 - \sT) \right]^\perp.
\end{equation}
\end{graybox}
This result was shown in~\cite{Wang_2017, Lee:2018eqa}. For the reader's convenience, we briefly review the derivation here.

We first note the basic identity:
\begin{equation}
    B(\sT a, b) = B(a, \sT b)^{-1}.
\end{equation}
This implies the following symmetry relation:
\begin{equation}
    B\bigl((1 + \sT)a, b\bigr) = B\bigl(a, (1 - \sT)b\bigr). 
\end{equation}
As a consequence, we obtain the inclusions:
\begin{equation}
    \Ker(1 - \sT) \subset \left[ \Ima(1 + \sT) \right]^\perp, \quad 
    \Ker(1 + \sT) \subset \left[ \Ima(1 - \sT) \right]^\perp. \label{eq:kernel-subset}
\end{equation}
Using the non-degeneracy of the bilinear form $B$, we obtain the inequalities:
\begin{equation}
    |\Ker(1 - \sT)| \le \frac{|\cA|}{|\Ima(1 + \sT)|}, \quad
    |\Ker(1 + \sT)| \le \frac{|\cA|}{|\Ima(1 - \sT)|}. \label{eq:kernel-bound}
\end{equation}
On the other hand, it is evident that
\begin{equation}
    |\cA / \Ker(1 + \sT)| = |\Ima(1 + \sT)|, \quad 
    |\cA / \Ker(1 - \sT)| = |\Ima(1 - \sT)|. \label{eq:image-vs-kernel}
\end{equation}
Combining equations~\eqref{eq:kernel-bound} and~\eqref{eq:image-vs-kernel}, we conclude that the inclusions in~\eqref{eq:kernel-subset} are in fact equalities:
\begin{equation}
    \Ker(1 - \sT) = \left[ \Ima(1 + \sT) \right]^\perp, \quad 
    \Ker(1 + \sT) = \left[ \Ima(1 - \sT) \right]^\perp. 
\end{equation}


 \bibliographystyle{ytamsalpha}
 \def\arxivfont{\rm}
 \baselineskip=.95\baselineskip
\bibliography{ref}

@artile{Lee:2018eqa,
    author = "Lee, Yasunori and Tachikawa, Yuji",
    title = "{A study of time reversal symmetry of abelian anyons}",
    eprint = "1805.02738",
    archivePrefix = "arXiv",
    primaryClass = "hep-th",
    reportNumber = "IPMU-18-0074",
    doi = "10.1007/JHEP07(2018)090",
    journal = "JHEP",
    volume = "07",
    pages = "090",
    year = "2018"
}

@article{Barkeshli:2016mew,
    author = "Barkeshli, Maissam and Bonderson, Parsa and Jian, Chao-Ming and Cheng, Meng and Walker, Kevin",
    title = "{Reflection and time reversal symmetry enriched topological phases of matter: path integrals, non-orientable manifolds, and anomalies}",
    eprint = "1612.07792",
    archivePrefix = "arXiv",
    primaryClass = "cond-mat.str-el",
    doi = "10.1007/s00220-019-03475-8",
    journal = "Commun. Math. Phys.",
    volume = "374",
    number = "2",
    pages = "1021--1124",
    year = "2019"
}

@article{Wang_2017,
   title={Anomaly Indicators for Time-Reversal Symmetric Topological Orders},
    eprint = "1610.04624",
    archivePrefix = "arXiv",
    primaryClass = "cond-mat.str-el",
   volume={119},
   ISSN={1079-7114},
   url={http://dx.doi.org/10.1103/PhysRevLett.119.136801},
   DOI={10.1103/physrevlett.119.136801},
   number={13},
   journal={Physical Review Letters},
   publisher={American Physical Society (APS)},
   author={Wang, Chenjie and Levin, Michael},
   pages={136801},
   year={2017},
   month=sep }

@article{Tachikawa_2020,
    author = "Tachikawa, Yuji",
    title = "{On gauging finite subgroups}",
    eprint = "1712.09542",
    archivePrefix = "arXiv",
    primaryClass = "hep-th",
    reportNumber = "IPMU-17-0183",
    doi = "10.21468/SciPostPhys.8.1.015",
    journal = "SciPost Phys.",
    volume = "8",
    number = "1",
    pages = "015",
    year = "2020"
}

@article{Benini_2019,
    author = "Benini, Francesco and C{\'o}rdova, Clay and Hsin, Po-Shen",
    title = "{On 2-Group Global Symmetries and their Anomalies}",
    eprint = "1803.09336",
    archivePrefix = "arXiv",
    primaryClass = "hep-th",
    reportNumber = "SISSA 10/2018/FISI, SISSA-10-2018-FISI",
    doi = "10.1007/JHEP03(2019)118",
    journal = "JHEP",
    volume = "03",
    pages = "118",
    year = "2019"
}

@article{Moore:1988qv,
    author = "Moore, Gregory W. and Seiberg, Nathan",
    title = "{Classical and Quantum Conformal Field Theory}",
    reportNumber = "IASSNS-HEP-88-39",
    doi = "10.1007/BF01238857",
    journal = "Commun. Math. Phys.",
    volume = "123",
    pages = "177",
    year = "1989"
}

@article{Orii_2025,
    author = "Orii, Ippo",
    title = "{On dimensions of (2+1)D abelian bosonic topological systems on unoriented manifolds}",
    eprint = "2502.13532",
    archivePrefix = "arXiv",
    primaryClass = "hep-th",
    doi = "10.1093/ptep/ptaf056",
    journal = "PTEP",
    volume = "2025",
    pages = "053",
    month = "2",
    year = "2025"
}

@article{Barkeshli_2019,
    author = "Barkeshli, Maissam and Bonderson, Parsa and Cheng, Meng and Wang, Zhenghan",
    title = "{Symmetry Fractionalization, Defects, and Gauging of Topological Phases}",
    eprint = "1410.4540",
    archivePrefix = "arXiv",
    primaryClass = "cond-mat.str-el",
    doi = "10.1103/PhysRevB.100.115147",
    journal = "Phys. Rev. B",
    volume = "100",
    number = "11",
    pages = "115147",
    year = "2019"
}

@inproceedings{Moore:1989vd,
    author = "Moore, Gregory W. and Seiberg, Nathan",
    title = "{Lectures on RCFT}",
    booktitle = "{1989 Banff NATO ASI: Physics, Geometry and Topology}",
    reportNumber = "RU-89-32, YCTP-P13-89",
    pages = "1--129",
    month = "9",
    year = "1989"
}

@article{Belov:2005ze,
    author = "Belov, Dmitriy and Moore, Gregory W.",
    title = "{Classification of Abelian spin Chern-Simons theories}",
    eprint = "hep-th/0505235",
    archivePrefix = "arXiv",
    month = "5",
    year = "2005"
}

@article{Barkeshli:2017rzd,
    author = "Barkeshli, Maissam and Cheng, Meng",
    title = "{Time-reversal and spatial-reflection symmetry localization anomalies in (2+1)-dimensional topological phases of matter}",
    eprint = "1706.09464",
    archivePrefix = "arXiv",
    primaryClass = "cond-mat.str-el",
    doi = "10.1103/PhysRevB.98.115129",
    journal = "Phys. Rev. B",
    volume = "98",
    number = "11",
    pages = "115129",
    year = "2018"
}

@article{Cui:2015cbf,
    author = "Cui, Shawn X. and Galindo, C{\'e}sar and Plavnik, Julia Yael and Wang, Zhenghan",
    title = "{On Gauging Symmetry of Modular Categories}",
    eprint = "1510.03475",
    archivePrefix = "arXiv",
    primaryClass = "math.QA",
    doi = "10.1007/s00220-016-2633-8",
    journal = "Commun. Math. Phys.",
    volume = "348",
    number = "3",
    pages = "1043--1064",
    year = "2016"
}

@misc{etingof2018reflectionfusioncategories,
      title={Reflection fusion categories}, 
      author={Pavel Etingof and César Galindo},
      year={2018},
      eprint={1803.05568},
      archivePrefix={arXiv},
      primaryClass={math.QA},
      url={https://arxiv.org/abs/1803.05568}, 
}

@article{WALL1963281,
title = {Quadratic forms on finite groups, and related topics},
journal = {Topology},
volume = {2},
number = {4},
pages = {281-298},
year = {1963},
issn = {0040-9383},
doi = {https://doi.org/10.1016/0040-9383(63)90012-0},
url = {https://www.sciencedirect.com/science/article/pii/0040938363900120},
author = {C.T.C. Wall}
}

@article{Nikulin1980,
  doi       = {10.1070/IM1980v014n01ABEH001060},
  url       = {https://dx.doi.org/10.1070/IM1980v014n01ABEH001060},
  year      = {1980},
  month     = feb,
  volume    = {14},
  number    = {1},
  pages     = {103--167},
  author    = {Nikulin, V. V.},
  title     = {{Integral symmetric bilinear forms and some of their applications}},
  journal   = {Mathematics of the USSR-Izvestiya}
}

@book{MacLane1995Homology,
  author    = {Saunders Mac Lane},
  title     = {Homology},
  series    = {Classics in Mathematics},
  publisher = {Springer Berlin, Heidelberg},
  year      = {1995},
  edition   = {1},
  pages     = {X + 422},
  isbn      = {978-3-540-58662-3},
  doi       = {10.1007/978-3-642-62029-4},
  note      = {Originally published as volume 114 in the series: Grundlehren der mathematischen Wissenschaften},
  issn      = {1431-0821},
  eissn     = {2512-5257},
  copyright = {Springer-Verlag Berlin Heidelberg},
  url       = {https://doi.org/10.1007/978-3-642-62029-4}
}

\end{document}